\documentclass[aip,pop,10pt,twocolumn]{revtex4-1}

\usepackage{graphicx}%
\usepackage{dcolumn}%
\usepackage{color}%
\usepackage{bm}%

\begin{document}

\title{Overstability of acoustic waves in strongly magnetized anisotropic MHD shear flows}

\author{E. S. Uchava}%
\affiliation{Faculty of Exact and Natural Sciences, Javakhishvili
Tbilisi State University, 3 Chavchavadze Ave., Tbilisi, 0179, Georgia}%
\affiliation{Abastumani Astrophysical Observatory, Ilia
State University, Tbilisi, Georgia}%
\affiliation{Nodia Institute of Geophysics, Javakhishvili Tbilisi
State University, Tbilisi, Georgia}

\author{B. M. Shergelashvili}%
\affiliation{Institut f\"ur Theoretische Physik IV: Weltraum- und
Astrophysik, Ruhr-Universitat Bochum, 44780 Bochum, Germany}%
\affiliation{Abastumani Astrophysical Observatory, Ilia
State University, Tbilisi, Georgia}%
\affiliation{CODeS, KU Leuven Campus Kortrijk, E. Sabbelaan 53, 8500
Kortrijk, Belgium}

\author{A. G. Tevzadze}%
\affiliation{Faculty of Exact and Natural Sciences, Javakhishvili
Tbilisi State University, 3 Chavchavadze Ave., Tbilisi, 0179, Georgia}%

\author{S. Poedts}%
\affiliation{Dept. of Mathematics, Centre for mathematical Plasma
Astrophysics, KU Leuven, Celestijnenlaan 200B, 3001 Leuven, Belgium.}%
\affiliation{Leuven Mathematical Modeling and Computational Science Center
(LMCC), KU Leuven, Celestijnenlaan 200B, 3001 Leuven, Belgium.}%

\begin{abstract}
We present a linear stability analysis of the perturbation modes in
anisotropic MHD flows with velocity shear and strong magnetic field.
Collisionless or weakly collisional plasma is described within the
16-momentum MHD fluid closure model, that takes into account not
only the effect of pressure anisotropy, but also the effect of
anisotropic heat fluxes. In this model the low frequency acoustic
wave is revealed into a standard acoustic mode and higher frequency
fast thermo-acoustic and lower frequency slow thermo-acoustic waves.
It is shown that thermo-acoustic waves become unstable and grow
exponentially when the heat flux parameter exceeds some critical
value. It seems that velocity shear makes thermo-acoustic waves
overstable even at subcritical heat flux parameters. Thus, when the
effect of heat fluxes is not profound acoustic waves will grow due
to the velocity shear, while at supercritical heat fluxes the flow
reveals compressible thermal instability. Anisotropic thermal
instability should be also important in astrophysical environments,
where it will limit the maximal value of magnetic field that a low
density ionized anisotropic flow can sustain.
\end{abstract}

\maketitle

\section{Introduction}

Physical properties of space plasmas in magnetized environments often reveal
their collisioness or weakly collisional character. Among these are the
magnetosphere of the Earth and other planets, solar and stellar winds,
astrophysical jets, magnetized accretion around compact objects, as well as
ionized low density interstellar and intercluster media. Such plasmas can not
be described by the simple one fluid magnetohydrodynamic (MHD) approach,
since they are characterized by pressure and temperature anisotropies with
respect to the magnetic field orientation.

Observations show strong anisotropy of solar wind plasmas
\cite{OMHC12,HWC12}. Similar features are observed in the near earth
magnetospheric plasmas. In this case, the mean free path is longer than the
gyroradius of the particle, and thus MHD description is not adequate. Similar
features should occur in the inner regions of accretion disks around black
holes, where the typical collision distance is longer then the event horizon
of the central black hole. In such situations, the magnetorotational
instability that is thought to drive accretion should be considered in the
collisionless limit \cite{Mikh08a,Mikh08b,QDH02,SQHS07,SQHS03}. It has been
also shown that pressure anisotropy has a significant effect on MHD
turbulence in the intracluster medium \cite{SdKFLN12,NdKS12,MS14}.

% onto CGL approximation
The most commonly used approach to analyze the anisotropic effects in the one
fluid approximation is the Chew-{Goldberger}-Low (CGL) approximation
\cite{CGL56}. This limit is often referred to as double adiabatic law MHD,
emphasizing the adiabatic equations of state that are the parallel and
perpendicular components of the pressure, respectively.

% On the importance of low frequency modes: failure neglecting heat fluxes
The CGL closure of the MHD equations does not include the effects of heat
fluxes on the flow. This in turn sets a high frequency limit to the
applicability of the model. Indeed, heat fluxes can be neglected for
processes with characteristic frequencies much higher than thermal buoyancy
effects. Hence, it can be marginally valid for the stability analysis, where
modes with zero or imaginary frequencies are involved. Indeed, the CGL model
was only a partial success when describing well known instabilities in
collisionless plasmas: the fire-hose and mirror instabilities. Although the
double adiabatic anisotropic description was able to resolve these
instabilities, the criteria of their onset were not properly replicated as
compared to the results of a more rigorous kinetic theory. As a matter of
fact, results of the CGL model, especially on the flow stability analysis,
should be considered in some sense as a somewhat crude estimate of the flow
stability. The deficiencies of the CGL model in the stability analysis of
anisotropic plasmas are already well known. Attempts have been made to
overcome them using linearized kinetic equations for the parallel and
transverse components of the pressure. The stability of low frequency modes
in this approach is studied based on the variational principle for the
potential energy \cite{Grigorev07,Grigorev08}.

% anisotropic MHD with heat fluxes
Still, probably the most effective method to overcome the deficiencies of the
CGL closure method in the MHD approximation is to consider three additional
moments during the closure method \cite{Oraevskii68}. In this limit, the
authors retain the heat fluxes along the magnetic field and derive a
so-called 16-momentum approximation of anisotropic MHD plasmas
\cite{Dzhalilov08,Kuznetsov09,Kuznetsov10,Dzhalilov10,Ramos03}. Indeed, the
16-momentum anisotropic MHD model proved to be successful in replicating
properties of classical instabilities (mirror, fire-hose) obtained in kinetic
theory using the fluid approximation \cite{Dzhalilov10,Kuznetsov09}. The
16-momentum MHD approximation has been used to describe the instability of
entropy waves in cosmic plasma \cite{Somov08}, or to propose a solar coronal
heating mechanism due to large scale wave mechanism \cite{Dzhalilov11}.

The purpose of the present paper is to analyze the linear stability of weakly
collisional plasmas in the presence of inhomogeneous background flow. Indeed,
velocity shear is a widely occurring factor in solar and stellar winds and
accretion flows. An attempt to describe the velocity shear effects in the
CGL-MHD approximation has shown a diversity of velocity shear effects
occurring in pressure anisotropic flows. Modifications to the fire-hose and
mirror instabilities have been analyzed \cite{Chagelishvili97}. Still, the
obtained results are to be considered within the limitations of the CGL
model.

In the present paper we adopt a simple shear configuration with a flow
aligned with the magnetic field, and a shear normal to the magnetic field
direction. We present the results of linear analysis of perturbation modes in
an anisotropic MHD shear flow in the strong magnetic field limit. In the
context of flow stability we focus on the low frequency solutions of the
dispersion equation. It seems that the existence of anisotropic heat fluxes
modifies the linear spectrum of the system. Stationary entropy modes become
compressional and we identify the fast and slow thermo-acoustic modes in the
linear spectrum. Hence, we study the stability of thermo-acoustic waves in
shear flows.

The mathematical formalism of the physical model is described in Sec. 2.
Here, we describe strongly magnetized anisotropic shear flow and introduce
linear perturbations. Sec. 3 discusses the linear spectrum of the problem in
the uniform and nonuniform flow limits. We study the linear spectrum and
instabilities occurring for different heat flux parameters. The results are
summarized in Sec. 4.

\section{Anisotropic MHD formalism}

In this section we formulate the 16-momentum MHD framework
\cite{Oraevskii68,Dzhalilov08} in order to study the linear stability of
compressible plasmas with anisotropic pressure and heat fluxes. The standard
MHD equations are complemented by anisotropic pressure terms as follows:
\begin{equation}
{\mathrm{d} \rho \over \mathrm{d} t} + \rho (\nabla \cdot \mathbf{V}) = 0 ~,
\end{equation}
$$
\rho{\mathrm{d} \mathbf{V} \over \mathrm{d} t} + \nabla \left(
P_{\perp} + {B^2 \over 8 \pi} \right) - {1\over 4\pi}
(\mathbf{B}\cdot \nabla) \mathbf{B} = \hskip 3cm
$$
\begin{equation}
~~~~ = (P_{\perp}-P_{\|}) \left[ \mathbf{h} (\nabla \cdot\mathbf{h})
+ (\mathbf{h}\cdot \nabla) \mathbf{h} \right] + \mathbf{h}
(\mathbf{h}\cdot \nabla) (P_{\perp}-P_{\|}) ~,
\end{equation}
\begin{equation}
{\mathrm{d} \mathbf{B} \over \mathrm{d} t} + \mathbf{B} (\nabla\cdot
\mathbf{V}) - (\mathbf{B} \cdot\nabla) \mathbf{V} = 0 ~,
\end{equation}
\begin{equation}
(\nabla \cdot\mathbf{B}) = 0 ~.
\end{equation}
Here, for the shortness of notations we use the convective derivative
$\mathrm{d}/\mathrm{d} t \equiv \partial / \partial t + (\mathbf{V} \cdot
\nabla)$, and $\mathbf{h} = \mathbf{B}/B$ is the unity vector field following
the streamlines of the magnetic field. Parallel and perpendicular pressure
($P_{\|}$,$P_{\perp}$) form two adiabatic invariants in the CGL model.
However, in the 16-momentum approximation the derivatives of these CGL
invariants are not zero due to the existence of nonzero heat fluxes
($S_{\|}$,$S_{\perp}$):
\begin{equation}
{\mathrm{d} \over \mathrm{d} t} \left( {P_{\|} B^{2} \over \rho^{3}}
\right) = -{B^2 \over \rho^3} \left[ B(\mathbf{h} \cdot\nabla)
{S_{\|}\over B} + {2 S_{\perp}\over B} (\mathbf{h}\cdot \nabla ) B
\right] ~,
\end{equation}
\begin{equation}
{\mathrm{d} \over \mathrm{d} t} \left( {P_{\perp} \over \rho B}
\right) = -{B \over \rho} (\mathbf{h}\cdot \nabla) {S_{\perp} \over
B^2 } ~.
\end{equation}
Finally, the 16-momentum closure model provides two more equations for the
heat fluxes:
\begin{equation}
{\mathrm{d} \over \mathrm{d} t } \left( {S_{\|} B^3 \over \rho^4}
\right) = -j {3 P_{\|} B^3 \over \rho^4} (\mathbf{h} \cdot\nabla)
{P_{\|} \over \rho} ~,
\end{equation}
\begin{equation}
{\mathrm{d} \over \mathrm{d} t} \left( {S_{\perp} \over \rho^2}
\right) = -j{P_{\|} \over \rho^2} \left[ (\mathbf{h}\cdot \nabla)
{P_{\perp} \over \rho} + {P_{\perp} \over \rho} {P_{\perp} - P_{\|}
\over P_{\|} B} (\mathbf{h}\cdot \nabla) B \right] ~.
\end{equation}
The parameter $j$ is used to switch to the CGL limit with zero heat
fluxes:
$$
j = \left\{
\begin{array}{lll}
1 & ~~\mathrm{when} & ( S_{\perp} \not= 0 ~, S_{\|} \not= 0 )  \\
0 & ~~\mathrm{when} & ( S_{\perp} = 0 ~, S_{\|} = 0 )
\end{array}
\right. ~,
$$
making it possible to set $S_{\perp}=S_{\|} \not= 0 $ consistently in
the Eqs. (7,8).

\begin{figure}[]
\begin{center}
\includegraphics[width = 6 cm]{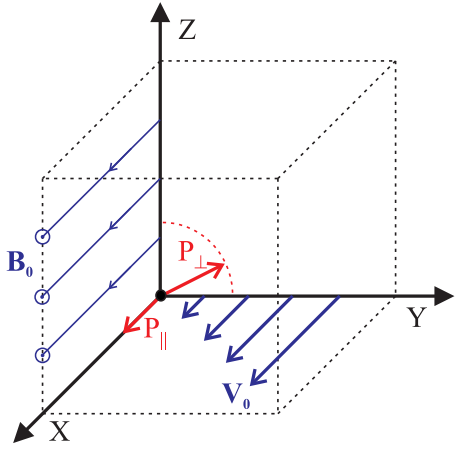}
\caption{Equilibrium MHD configuration with shear flow along the
background magnetic field. Direction parallel to the magnetic field
is set by $X$ axis, while the perpendicular direction lays in the
$YZ$ plane. The flow along the $X$ axis is nonuniform in the $Y$ direction.}
\label{MHD_config}
\end{center}
\end{figure}

\subsection{Background Flow}

We consider a stationary shear flow along the magnetic field with
constant background density and temperature:
\begin{equation}
\mathbf{V_0} = (Ay, 0, 0 ) ~,~~ \mathbf{B_0} = (B_0, 0, 0 ) ~,
\end{equation}
where $A$ is the velocity shear rate. Fig. \ref{MHD_config} schematically
shows the flow configuration.

We define an anisotropy parameter $\alpha$ as the ratio between the
perpendicular and parallel components of the background pressure or heat
fluxes:
\begin{equation}
\alpha \equiv {P_{\perp 0} / P_{\| 0}} = {S_{\perp 0} / S_{\| 0}} ~.
\end{equation}
We may introduce sound speeds parallel and perpendicular to the
background flow:
\begin{equation}
C_{\|}^2 \equiv P_{\| 0}/\rho_0 ~,~~ C_{\perp}^2 \equiv P_{\perp
0}/\rho_0 ~,
\end{equation}
Hence we may define a non-dimensional anisotropic heat flux
parameter as follows:
\begin{equation}
\gamma \equiv {S_{\| 0} \over P_{\| 0} C_{\|}} ~.
\end{equation}
The parameter $\gamma$ is the normalized measure of the heat flux
effects in pressure anisotropic MHD flows. When the heat fluxes can
be neglected ($j=0$, $\gamma=0$) we obtain the double adiabatic
limit of the CGL anisotropic model.

\subsection{Linear Perturbations}

To proceed with a linear analysis of the anisotropic MHD shear flow we split
the physical variables into background and perturbation components:
\begin{eqnarray}
{\bf \rho} &=& {\bf \rho_0} + {\bf \rho^\prime} ~,~~ \nonumber \\
{\bf V} &=& {\bf V_0} + {\bf V^\prime} ~, \nonumber \\
{\bf B} &=& {\bf B_0} + {\bf B^\prime} ~, \\
P_\parallel &=& P_{0 \parallel} + P_{\parallel}^\prime ~,~~~
P_\perp = P_{0 \perp} + P_{\perp}^\prime ~, \nonumber \\
S_\parallel &=& S_{0 \parallel} + S_{\parallel}^\prime ~,~~~ S_\perp
= S_{0 \perp} + S_{\perp}^\prime ~. \nonumber
\end{eqnarray}
Hence, we may employ the shearing sheet formalism to analyze the linear
dynamics of perturbations in shear flow \cite{Goldreich65}. In this limit the
shearing transformation is used to transfer the spatial dependence due to the
velocity shear into a temporal variation of the wave-numbers in the direction
of the velocity shear and analyze the initial value problem. This method,
often called a non-modal approach, is based on the study of the evolution of
the spatial Fourier harmonics (SFH) in time
\cite{Chagelishvili97,ChChLT97,T98}. Hence, we follow the non-modal analysis
and introduce the spatial Fourier transformation with time dependent
wave-numbers:
\begin{equation}
\Psi({\bf r},t) \propto \psi({\bf K},t) \exp \left({\rm i} K_x x +
{\rm i} K_y(t) y + {\rm i} K_z z \right) ~, \label{SFH}
\end{equation}
where
\begin{equation}
K_y(t) = K_{y0} - A K_x t ~.
\end{equation}
Here $\Psi({\bf r},t)$ and $\psi({\bf K},t)$ are generalized vectors
introduced for shortness of notations:
\begin{equation}
\Psi \equiv \left( {\rho^\prime \over \rho_0} ,~ {P_{\|}^\prime
\over P_{\| 0}} ,~ {P_{\perp}^\prime \over P_{\perp 0}} ,~
{S_{\|}^\prime \over P_{\| 0}} ,~ {S_{\perp}^\prime \over P_{\perp
0}} ,~ {{\bf V^\prime}\over V_A} ,~ {{\bf B^\prime} \over B_0}
\right) ,
\end{equation}
\begin{equation}
\psi \equiv \left( {\rm i} {\varrho} ,~ {\rm i} {p_{\|} } ,~ {\rm i}
{p_{\perp}} ,~ {\rm i } {s_{\|}} ,~ {\rm i } {s_{\perp}} ,~ {\bf v}
,~ {\rm i} {\bf b} \right) ~.
\end{equation}
Note that the perturbation SFH $\psi$ is introduced in the non-dimensional
form with complex coefficients used to account for intrinsic phase
differences between the kinetic and thermodynamic quantities in the
wave-number space.

Linearized with respect to assumed to be small perturbations, the PDE system
(1-8) can be transformed into a system of ODEs for the spatial harmonics of
perturbations using the Fourier expansion (\ref{SFH}). Hence, the system
describing the evolution of linear perturbations in time will read as
follows:
\begin{eqnarray}
\dot{\varrho} &=& -k_x v_x - k_y v_y - k_z v_z
\\
\dot{v}_x &=& -R v_y + \beta_{\|} k_x p_{\|} +
(\beta_\perp-\beta_\|) k_x b_x
\\
\dot{v}_y &=& \beta_\perp k_y p_\perp - (1+\beta_\perp-\beta_\|) k_x
b_y + k_y b_x
\\
\dot{v_z} &=& \beta_\perp k_z p_\perp - (1+\beta_\perp-\beta_\|) k_x
b_z + k_z b_x
\\
\dot{p_\|} &=& -3 k_x v_x- k_y v_y- k_z v_z - ik_x s_\| +
\\
&& ~~~~~~~~~~~~~~~~~ + {\rm i} \gamma (1-2\alpha^2)\beta_\|^{1\over2} k_x
b_x - 2{\rm } R b_y ~, \nonumber
\\
\dot{p_\perp} &=& - k_x v_x - 2 k_y v_y - 2 k_z v_z  - i k_x s_\perp
+
\\
&& ~~~~~~~~~~~~~~~~~~~~~~~~~~~  + 2{\rm i} \gamma \beta_{\perp}^{1\over2}
k_x b_x + {\rm } R
b_y ~,  \nonumber \\
 \dot{s_\|} &=& -4 \gamma \beta_\|^{1\over2}k_x v_x -
 \gamma \beta_\|^{1\over2}k_y v_y - \gamma \beta_\|^{1\over2}k_z v_z + \\
&& ~~~~~~~~~~~~~~~~~  + 3{\rm i}j \beta_\|k_x \left( \rho - p_\|
\right) - 3 \gamma R \beta_\|^{1\over2} b_y ~, \nonumber \\
\dot{s_\perp} &=& - 2 \gamma \beta_{\perp}^{1\over2}k_x v_x -
2\gamma\beta_{\perp}^{1\over2}
k_yv_y - 2 \gamma\beta_{\perp}^{1\over2} k_z v_z + \\
&& ~~~~~~~~~~~~~~~~~ + {\rm i} j \beta_\| k_x \left( \rho - p_\perp
+
(1-\alpha) b_x \right) ~, \nonumber \\
\dot{b_y} &=& k_x v_y \\
\dot{b_z} &=& k_x v_z \\
0 &=& k_x b_x + k_y b_y + k_z b_z,
\end{eqnarray}
where $\textbf{k} \equiv \textbf{K} / K_\perp$ is non-dimensional wave-vector
with $K_\perp = (K_y^2+K_z^2)^{1\over2}$, $\tau \equiv V_A K_\perp t$ is
non-dimensional time, $\dot{\psi}$ denotes time derivative of $\psi$ with
respect to $\tau$. We define longitudinal and transverse plasma $\beta$
parameters:
$$
\beta_{\|} \equiv {C_{\|}^2 \over V_A^2} ~,~~~ \beta_{\perp} \equiv
{C_{\perp}^2 \over V_A^2} ~,
$$
and $R \equiv A/ (V_A K_{\perp})$ is the non-dimensional shear rate.

\section{Linear Spectrum}

Strictly speaking, the linear spectrum of the problem can be obtained using
the Fourier expansion of the spatial harmonics of the perturbations in time.
In fact, the coefficients of the Eqs. (18-28) are explicitly time dependent
due to the velocity shear of the flow. However, it is still possible to
employ the adiabatic approximation when
\begin{equation}
 \dot{\psi}({\rm k},\tau) \approx - {\rm i} \omega
\psi({\rm k},\tau) ~,
\end{equation}
where $\omega$ can be a slowly varying function in time. The
considered approximation is valid for low or moderate velocity shear
of the flow and is strongly justified for low frequency modes of the
spectrum. The linear spectrum of anisotropic MHD uniform flow (with
zero shear) is analyzed in \cite{Dzhalilov08}. Velocity shear
introduces farther complications to the general dispersion equation
(see Eqs. A1-A4 in the Appendix). In the present paper we intend to
describe the compressible sound waves in more detail using the cold
plasma approximation. In this limit the magnetic pressure dominates
over the hydrodynamic one and hence:
\begin{equation}
\beta_\|,\beta_{\perp} \ll 1 ~.
\end{equation}
This leads to a somewhat simplified dispersion equation for the considered
problem, that can be factorized into the high frequency
\begin{eqnarray}
&\omega^2 - k_x^2 = 0
\end{eqnarray}
and low frequency modes:
\begin{equation}
(\omega^2-j\beta_\|k_x^2)  D_0 -{\rm i} R \beta_\perp {k_x k_y \over
k^2} D_1 = 0 ~,
\end{equation}
where
\begin{eqnarray}
D_0 = \omega^4 - 3(j+1) \beta_\| k_x^2 \omega^2 - 4\gamma
\beta_\|^{3\over2} k_x^3 \omega + 3 j \beta_\|^2 k_x^4 ~, ~~~~
\end{eqnarray}
\begin{eqnarray}
D_1 &=& (2\omega^5+2\alpha \gamma \beta_\|^{1\over2}k_x
\omega^4-(5+7j)
\beta_\|k_x^2\omega^3 -  ~~~~~~~~ \nonumber \\
&-& (2\alpha(2+3j)+7)\gamma \beta_\|^{3\over2}k_x^3\omega^2 + \nonumber \\
&+& (j(3j+5)-6\alpha \gamma^2)\beta_\|^2k_x^4\omega + \nonumber \\
&+& 3\gamma j \beta_\|^{5\over2}k_x^5) ~.
\end{eqnarray}

\begin{figure}[]
\begin{center}
\includegraphics[width = 6 cm ]{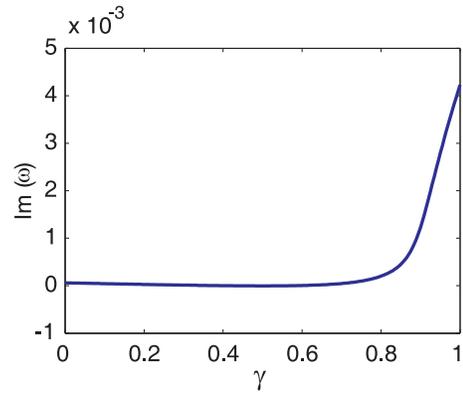}
\caption{Growth rate of the compressible heat flux instability vs
heat flux parameter $\gamma$ in the uniform flow: $R=0$. Figure
reveals the critical value of heat flux parameter $\gamma_{cr} =
0.85$ that is necessary for the instability.} \label{gamma_0}
\end{center}
\end{figure}

We can solve the dispersion equation in the zero shear limit when it reduces
to $(\omega^2-j\beta_\|k_x^2) D_0 = 0$. In the absence of heat fluxes
($j=\gamma=0$), the dispersion relation reduces to $\omega^2(\omega^2 -
3\beta_\|k_x^2) = 0$ with the well known CGL solution for acoustic waves:
$\omega^2 = 3\beta_\|k_x^2$. In case of nonzero heat fluxes we get three
different periodic solutions:
\begin{itemize}
\item fast thermo-acoustic mode with $\omega_{+} =
      \beta_\|^{1\over2} k_x \eta_{+}$,
\item acoustic mode with $\omega_s^2 = \beta_\|k_x^2$,
\item slow thermo-acoustic mode with $\omega_{-} =
      \beta_\|^{1\over2} k_x \eta_{-}$,
\end{itemize}
where $\eta_\pm$ are the upper and lower pair solutions of the equation:
$\eta^4-6\eta^2-4\gamma\eta+3=0$. We identify these solutions as fast and
slow thermo-acoustic modes since they obey the property: $\eta_{+}^2 > 1 >
\eta_{-}^2$. The linear instability for uniform flow with zero shear of
velocity is found when heat flux parameter exceeds some critical value.
Indeed, Fig. \ref{gamma_0} shows the growth rate of the slow thermo-acoustic
mode that becomes unstable at supercritical heat flux rates:
$\gamma>\gamma_{cr} \approx 0.85$.

\begin{figure}[]
\begin{center}
\includegraphics[width = 7.5 cm]{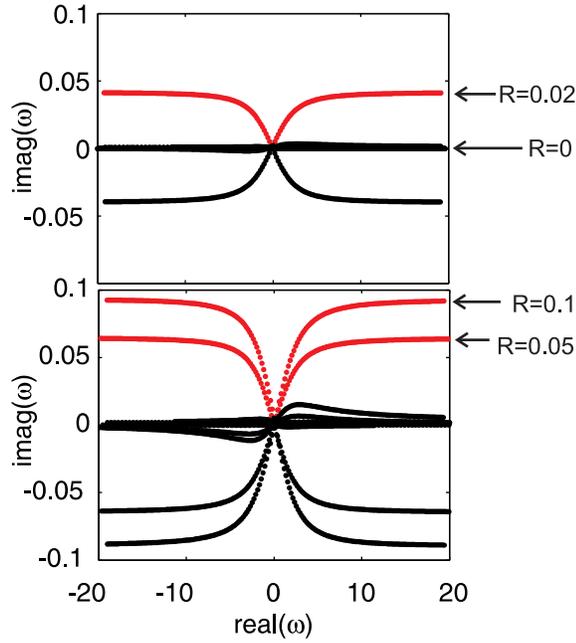}
\caption{Dispersion curves at subcritical heat flux parameter
$\gamma=0.5$, anisotropy parameter $\alpha=2$ and different values
of normalized shear parameter: $R =0, 0.02, 0.05, 0.1$. Here
$\beta_\| = 0.01$ and harmonics with $k_y/k_x=2$ are shown for
different values of $k_z/k_x$. Instability growth rate grows with
the velocity shear of the flow.} \label{gamma_05}
\end{center}
\end{figure}

\begin{figure}[]
\begin{center}
\includegraphics[width = 7.5 cm]{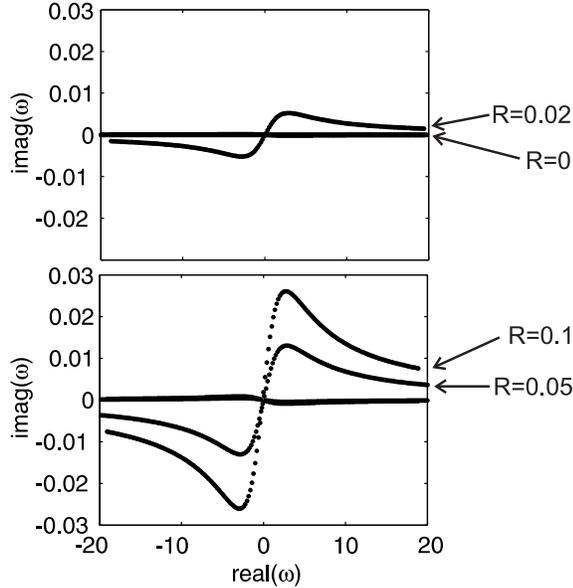}
\caption{Dispersion curves at supercritical heat flux parameter
$\gamma=1.5$, anisotropy parameter $\alpha=2$ and different values
of normalized shear parameter: $R =0, 0.02, 0.05, 0.1$. Here
$\beta_\| = 0.01$ and harmonics with $k_y/k_x=2$ are shown for
different values of $k_z/k_x$. Effect of the velocity shear is
dominated by the heat flux instability growth rate (see Fig.
\ref{gamma_15_0}).} \label{gamma_15}
\end{center}
\end{figure}

\begin{figure}[h]
\begin{center}
\includegraphics[width = 6 cm]{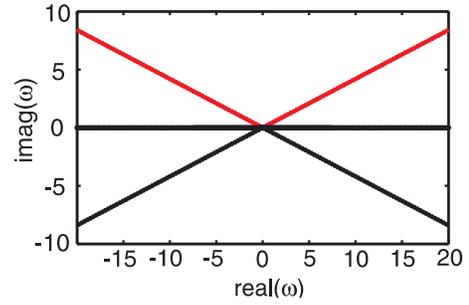}
\caption{Dispersion curves at supercritical heat flux parameter
$\gamma=1.5$ in zero shear limit: $R=0$. Figure shows that the
growth rate of exponential thermal instability is much higher
compared to the growth rate of acoustic overstability at the same
parameters of the system (see Fig. 4). } \label{gamma_15_0}
\end{center}
\end{figure}

\begin{figure}[]
\begin{center}
\includegraphics[width = 6 cm]{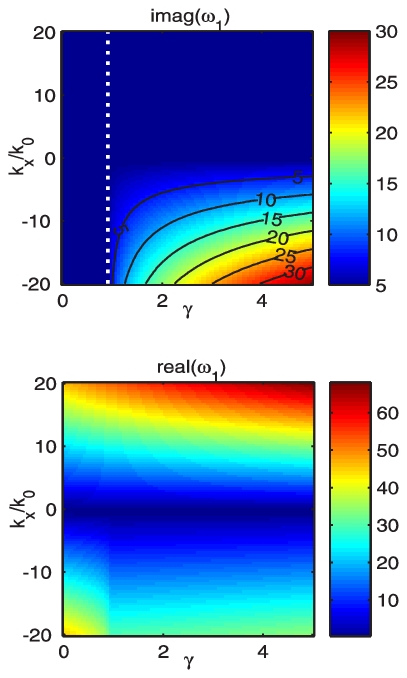}
\caption{Surfaces of the growth rate ($Im(\omega)$) and frequency
($Re(\omega)$) of the fast thermo-acoustic mode are shown for the
different values of stream-wise wave-number $k_x$ and heat flux
parameter $\gamma$. Vertical white dashed lines shows the critical
value of $\gamma$ parameter for the instability due to heat fluxes.
Complex part reveals the instability area, where the real part of
the solution is not vanishing. Moreover, in this area real part is
similar to the complex one, thus identifying the overstability of
the mode. The growth rates of the overstability increase with the
growth of heat flux parameter and streamwise wavenumber.}
\label{surf_1}
\end{center}
\end{figure}

\subsection{Shear flow solutions}

In the presence of velocity shear the linear modes are modified proportional
to the transverse plasma beta $\beta_\perp$ and velocity shear rate $R$, see
the second term in Eq. (32). For more specific results we use numerical
methods to calculate the solutions of the dispersion equation.

Fig. \ref{gamma_05} shows the solution of the dispersion equation for
different velocity shear parameters. Numerical solutions of the dispersion
equation (32) show the growth rate of the thermo-acoustic mode. The
instability is stronger for higher velocity shear rates. The growth rates
reach maximum values for the perturbations with the smallest stream-wise
length-scales ($k_x \gg 1$).

Dispersion curves at the supercritical heat flux parameter
($\gamma>\gamma_{cr}$) and zero velocity shear ($R=0$) are shown in Fig.
\ref{gamma_15_0}. This instability is nearly insensitive to the velocity
shear rate. The heat flux instability is stronger compared to the velocity
shear effects shown on Fig. \ref{gamma_15}. Thus, at supercritical values and
any value of the velocity shear rate, the plasma will reveal strong
compressible thermal instability features.

Figs. \ref{surf_1} shows the frequency and the growth rate of slow
thermo-acoustic waves for different stream-wise wave-numbers and heat flux
parameters. Numerical results show that the real an complex parts of the
solutions are of the same order in the area of instability. Thus, energy
growth is achieved through thermo-acoustic wave overstability: the time-scale
of the growth is comparable with the oscillation period of the growing wave
mode. In this case, in the vicinity of the outflow acoustic waves will grow
due to the shear flow overstability mechanism. This will lead to an enhanced
dissipation of compressible waves and flow heating. At larger distances,
where $r>r_{\rm cr}$ and $\gamma > \gamma_{\rm cr}$ an exponential
instability will develop due to anisotropic heat flux instabilities. This in
turn can lead to a fragmentation into magnetized clouds at large scales.

\section{Summary}

We presented the linear stability analysis of an anisotropic strongly
magnetized MHD shear flow within the 16-momentum MHD approximation. We
identify three compressible solutions of the dispersion equation as fast and
slow thermo-acoustic and standard acoustic wave modes in strongly magnetized
anisotropic collisionless plasmas. We find the critical value of the
normalized heat flux parameter that leads to compressible thermal instability
in uniform flows with strong magnetic field: $\gamma_{\rm cr} = 0.85$.

We study the effects of velocity shear on the thermo-acoustic waves. It seems
that background flow inhomogeneity leads to the overstability of
thermo-acoustic modes. In this case, compressible perturbations in  the
inhomogeneous anisotropic plasma outflow will grow in amplitude and lead to
enhanced dissipation and heating. The effect of anisotropic heat fluxes is
most profound at supercritical heat flux parameters, when an exponential
instability develops. In this regime compressible shear flow overstability is
well dominated by exponential thermal heat flux instability.

The overstability of acoustic waves described in the present paper can have
important consequences for the space plasma dynamics. This process will draw
shear flow energy into the compressible waves and eventually to heat via
dissipation in solar and stellar winds, leading to heating due to intrinsic
wind shear.

A better understanding of the turbulent processes responsible for the heating
and acceleration of the solar and stellar winds is of special interest here.
In this respect, overstability of acoustic waves due to anisotropic heat
fluxes and velocity shear can contribute to the in-situ heating of the
astrophysical winds that are observed to be hotter than predicted by a simple
adiabatic expansion model. On the other hand, it is interesting to examine
the role of the additional modes appearing in the linear spectrum in the
turbulent heating models like recently reported \cite{shergelashvili12}.
Another important aspect of the developed model is to understand the role of
the thermal modification of the MHD wave modes found within the 16-momentum
approximation. It is known that shear flow induces wave couplings and
enhanced dissipation processes e.g., see the self-heating mechanism described
in the solar coronal heating context  \cite{shergelashvili2006}.

Many ionized flows from astrophysical objects fall into the category of
expanding anisotropic plasmas. Among these are solar and stellar winds,
magnetized outflows from galaxies or even spherically expanding shells of
supernovae. It seems that the radial decay laws of expanding outflows may
define their local thermal stability far away from the source. Indeed, if the
heat flux parameter increases with outflow, at some distance from the source
it may exceed the critical value ($\gamma>\gamma_{cr}$) and the exponential
instability will develop. Such a situation can occur not only in strongly
magnetized stellar winds, but also in specific types of jet outflows. Similar
arguments can be applied to the opposite case of accreting flows, when
anisotropic plasma falls onto a central massive object. In this case the
thermal instability can be developed during the accretion process.

The phenomena described in the present paper may be also important in
extremely rarified plasmas, such as intracluster gas or galactic winds. The
low density of these outflows provides an environment where the local Larmor
radius of ions is shorter than mean free path of the particles. Often these
outflows are strongly magnetized and exhibit nonuniform velocity features.
Galactic winds are thought to carry dynamo generated strong magnetic fields
at larger scales, where they are observed. In such situations, the local
exponential thermal instability will lead to the destruction of a directed
flow and the buoyant generation of magnetic bubbles in the outflow. This
mechanism will also limit the maximal value of magnetic field that such
rarified ionized flow can sustain.

\section*{Acknowledgements}

The research leading to these results has partially received funding from the
European Community's Seventh Framework Programme (FP7/2007-2013), within the
framework of SOLSPANET project under grant agreement
FP7-PEOPLE-2010-IRSES-269299. Authors appreciate useful discusions with Prof.
Namig Dzhalilov. E. Uchava would like to acknowledge hospitality of the
KULeuven during her visits in Kortrijk.

\appendix \section*{Appendix}
\setcounter{section}{1} \setcounter{equation}{0}

The linear dispersion equation of the anisotropic MHD shear flow with heat
fluxes can be obtained from Eqs. (18-28) using adiabatic approximation set by
Eq. (29). Hence, 10th order system yields dispersion equation that can be
factorized into the following two equations:
\begin{equation}
\omega^2-(1+(\alpha-1)\beta_{\|})k_x^2 = 0 ~,
\end{equation}
\begin{equation}
\omega^8 + \sum_{n=0}^6{ (\beta_{\|}^{1/ 2} k_x)^{6-n}
\left[ a_n \omega^n + {\rm i} R k_y b_n \right] } = 0 ~,
\end{equation}
where $k_\perp \equiv k_x^2 + k_y^2$,
\begin{equation}
\begin{array}{lll}
a_0 &=& 3 j \left(k^2 - (1-\alpha)\beta_\| (k_x^2-2 \alpha
k_\perp^2) \right) , \\
%$$
%$$
a_1 &=& 4j \gamma
\left(((1-\alpha)\beta_\|-1)k_x^2 +
(\alpha(2\alpha^{3\over2}-{\alpha \over
2}(\alpha^{1\over2}-1)-\right.\\
&& ~~~~~~~~~~~~~~~~\left. 2+\alpha)\beta_\| - 1)k_\perp^2 \right), \\
%$$
%$$
a_2 &=&
\left((10\alpha-18)j-4\alpha^{3\over2} \gamma^2\right)\alpha
\beta_\| k_\perp^2 + 3(2-3\alpha)j\beta_\| k_x^2 \\
&& ~~~~~~~~~~~~~~~~~~~~~~~~~~~~~~~~~~~~~~~~~~~~~~~ - 9jk^2 , \\
%$$
%$$
a_3 &=&
4\gamma((1+(\alpha-1+j)\beta_\|)k_x^2 +
(1+\alpha\left(2-\alpha\right. \\
&&
~~~~~~~~~~~~~~~~~~~~~~~~\left.{\alpha^{3\over2}+1\over2}\right)\beta_\|))
k_\perp^2,\\
%$$
%$$
a_4 &=& (4j+3)k^2 + ((8-\alpha)j + 6-\alpha)\alpha \beta_\|
k_\perp^2 \\
&& ~~~~~~~~~~~~~~~~~~~~~~~~ + ((5+4\alpha)j+3\alpha-3)\beta_\| k_x^2
~,
\\
a_5 &=& -4 \gamma \beta_\| k_x^2 , \\
a_6 &=& -(1+2 \alpha \beta_\|)k^2 - (4j +2 -\alpha) \beta_\| k_x^2 ,
\end{array}
\end{equation}
and
\begin{equation}
\begin{array}{rllll}
b_0 &=& 3j \gamma \alpha \beta_{\|}^{1/2} ~,
&b_1 =&{ (8j-6\alpha^{1\over2} \gamma^2) \alpha \beta_{\|}^{1/2} }~, \\
b_2 &=& -((4+6j)\alpha^{1\over2}+7)
\gamma \alpha \beta_{\|}^{1/2} ~,
&b_3 =& -(7j+5) \alpha \beta_{\|}^{1/2} ~, \\
b_4 &=& 2\gamma \alpha^{3\over2}
\beta_{\|}^{1/2} ~,
&b_5 =& 2 \alpha \beta_{\|}^{1/2} ~, \\
b_6 &=& 0 ~. &&
\end{array}
\end{equation}
The Eq. (A1) shows solution for the well known fire-hose mode, while the
solutions of the Eq. (A2) identify the remaining modes that are affected by
the presence of the velocity shear ($R \not=0$).

\end{document}